\providecommand{\FIGONE}{
\begin{figure}[ht]
  \includegraphics[width=7truecm]{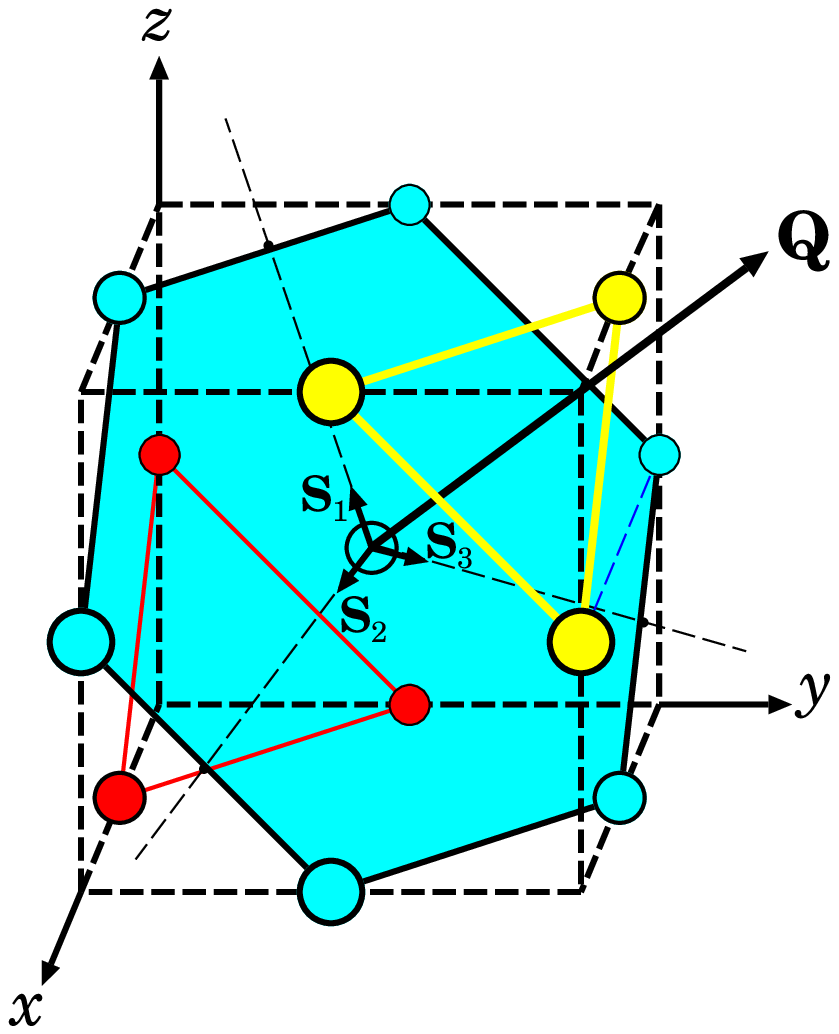}
\caption{
Sketch of a NiO crystal (only Ni atoms are shown).  Wave vector ${\bf Q} =
 (\frac{1}{2} \frac{1}{2} \frac{1}{2})$ characterizes an antiferromagnetic
 modulation direction.  Three arrows denoted as ${\bf S}_1$, ${\bf S}_2$,
 ${\bf S}_3$, in a plane perpendicular to ${\bf Q}$ indicate possible spin
 directions ($S$ domains).  Spins for solid, gray, and open circles align
 alternately. 
\label{fig.cryst}}
\end{figure}
}

\providecommand{\TABLEI}{
\begin{table}[h]
\caption{\label{tab.1} Calculated band gap and magnetic moments, compared
 with those in the experiments. $2S$ and $L$ are spin and orbital moments,
 respectively.  In the LDA$+U$ calculations, the effective parameters of (1)
 $U_{\rm eff}=2$ eV and (2) $U_{\rm eff}=5$ eV are employed.}
\begin{ruledtabular}
\begin{tabular}{lccccc}
            &  $E_{\rm gap}$ [eV] 
            &  $2S$   [$\hbar$]  &   $L$  [$\hbar$]
            &  $L+2S$ [$\hbar$]  & $L/S$    \\
\hline
 LDA        & 0.408    &   1.210   &   0.151   &   1.361      &   0.25    \\
 LDA$+U$(1) & 1.438    &   1.467   &   0.187   &   1.654      &   0.25    \\
 LDA$+U$(2) & 2.839    &   1.647   &   0.233   &   1.880      &   0.28    \\
 Exp.\footnotemark[1]
            & 4.3 \footnotemark[2]
            & 1.88     & 0.32    & 2.20, 1.90\footnotemark[3] &   0.34    \\
\end{tabular}
\end{ruledtabular}
  \footnotetext[1]{Unless noted, Ref.~\onlinecite{Fernandez}.}
  \footnotetext[2]{Reference~\onlinecite{Sawatzky}. }
  \footnotetext[3]{Reference~\onlinecite{Cheetham}. }
\end{table}
}
\providecommand{\FIGTWO}{
\begin{figure}[ht]
\includegraphics[width=7truecm]{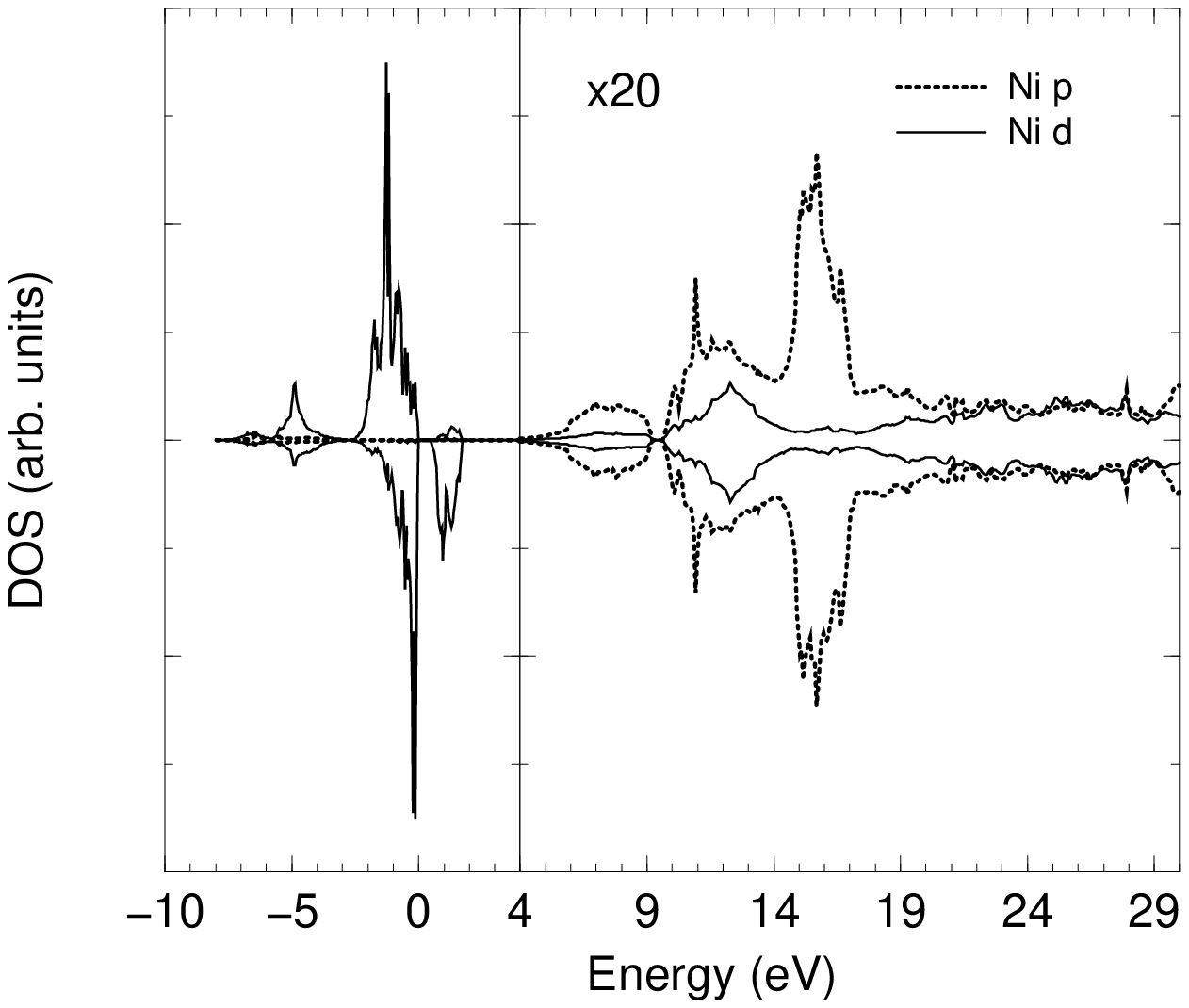}
\caption{Projected density of states of the Ni $3d$ and $4p$ states
 calculated in the LDA.  All energies are given with respect to the top of
 the valence band.  \label{fig.dos}}
\end{figure}
}

\providecommand{\FIGTHREE}{
\begin{figure}[ht]
\includegraphics[width=7truecm]{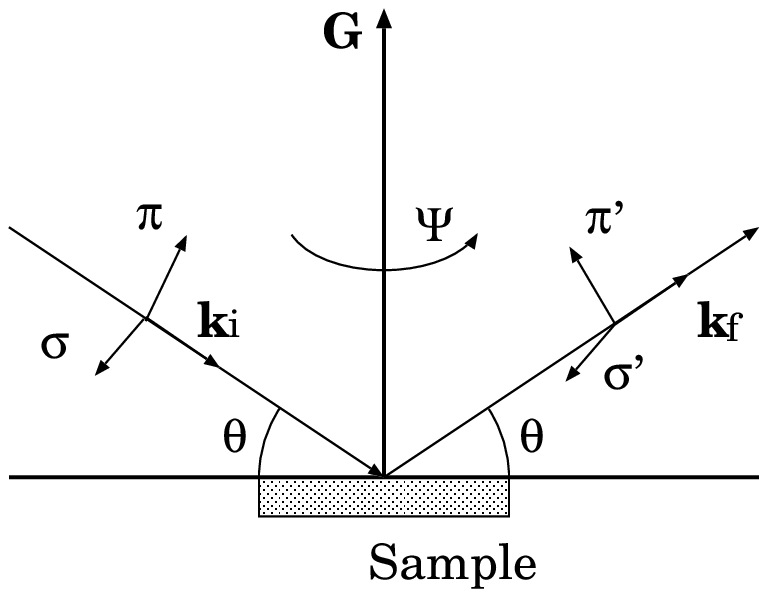}
\caption{ Geometry of x-ray scattering. Incident photon with wave
 vector ${\bf k}_i$ and polarization $\sigma$ or $\pi$ is scattered into the
 state with wave vector ${\bf k}_f$ and polarization $\sigma'$ or $\pi'$ at
 Bragg angle $\theta$.  The sample crystal is rotated by azimuthal angle
 $\psi$ around scattering vector ${\bf G}={\bf k}_f-{\bf k}_i$.
  \label{fig.geom}}
\end{figure}
}

\providecommand{\FIGFOUR}{
\begin{figure}[ht]
\includegraphics[width=7truecm]{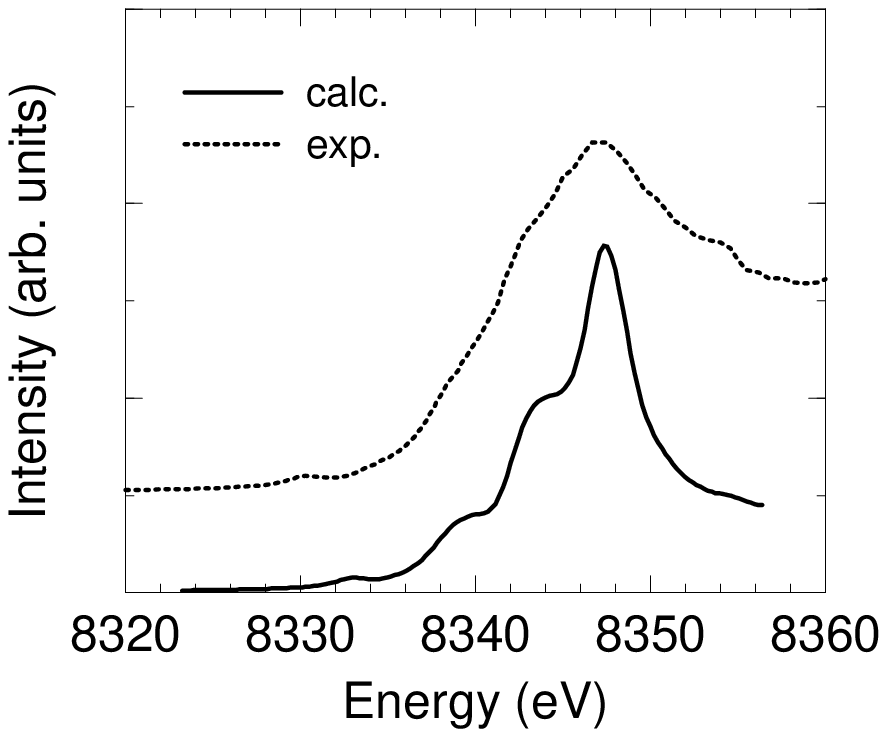}
\caption{ Absorption coefficient $A(\omega)$ around the $K$ edge and
 the experimental fluorescence spectrum (Ref.~\onlinecite{Neubeck2}).  
 The core-hole energy was adjusted such that the peak position coincided
 with that in the experiment. \label{fig.absorp}}
\end{figure}
}

\providecommand{\FIGFIVE}{
\begin{figure}[ht]
\includegraphics[width=8truecm]{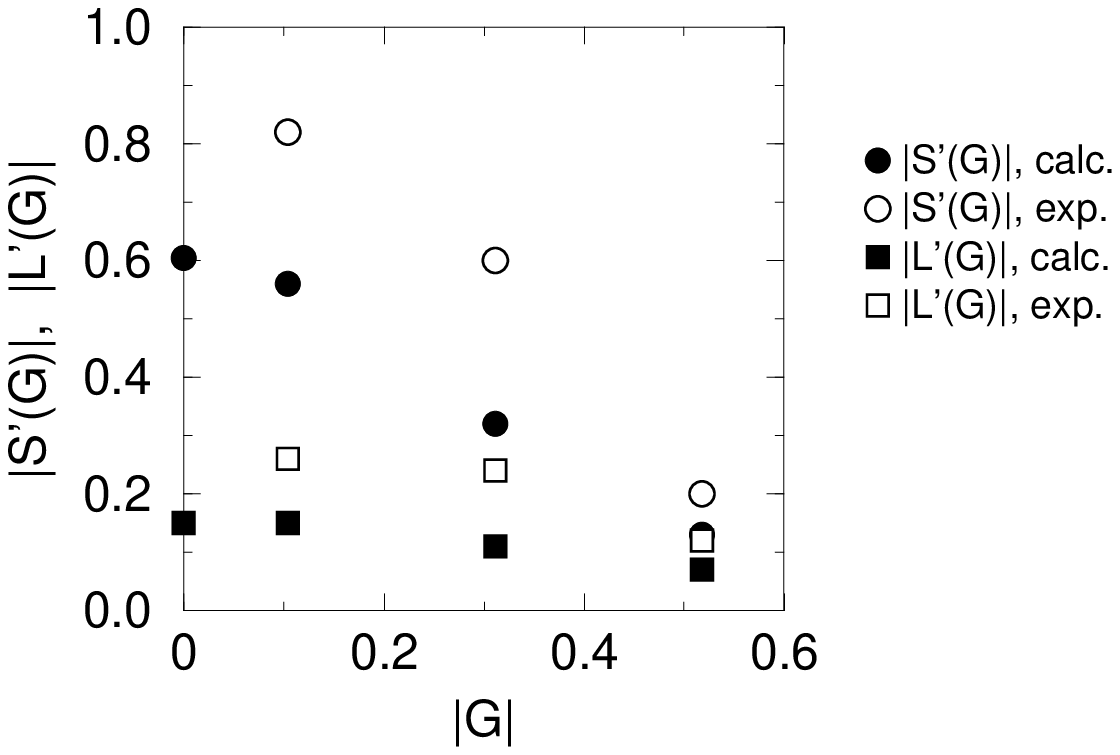}
\caption{ Form factors of the spin and orbital moments.  Circles: the spin
 moment per Ni site, $|{\bf S'}({\bf G})|\equiv |{\bf S}({\bf
 G})|/\sqrt{N}$.  Squares: the orbital moment per Ni site, $|{\bf L'}({\bf
 G})|\equiv |{\bf L}({\bf G})|/\sqrt{N}$.  \label{fig.form}}
\end{figure}
}

\providecommand{\FIGSIX}{
\begin{figure}[ht]
\includegraphics[width=.72\linewidth]{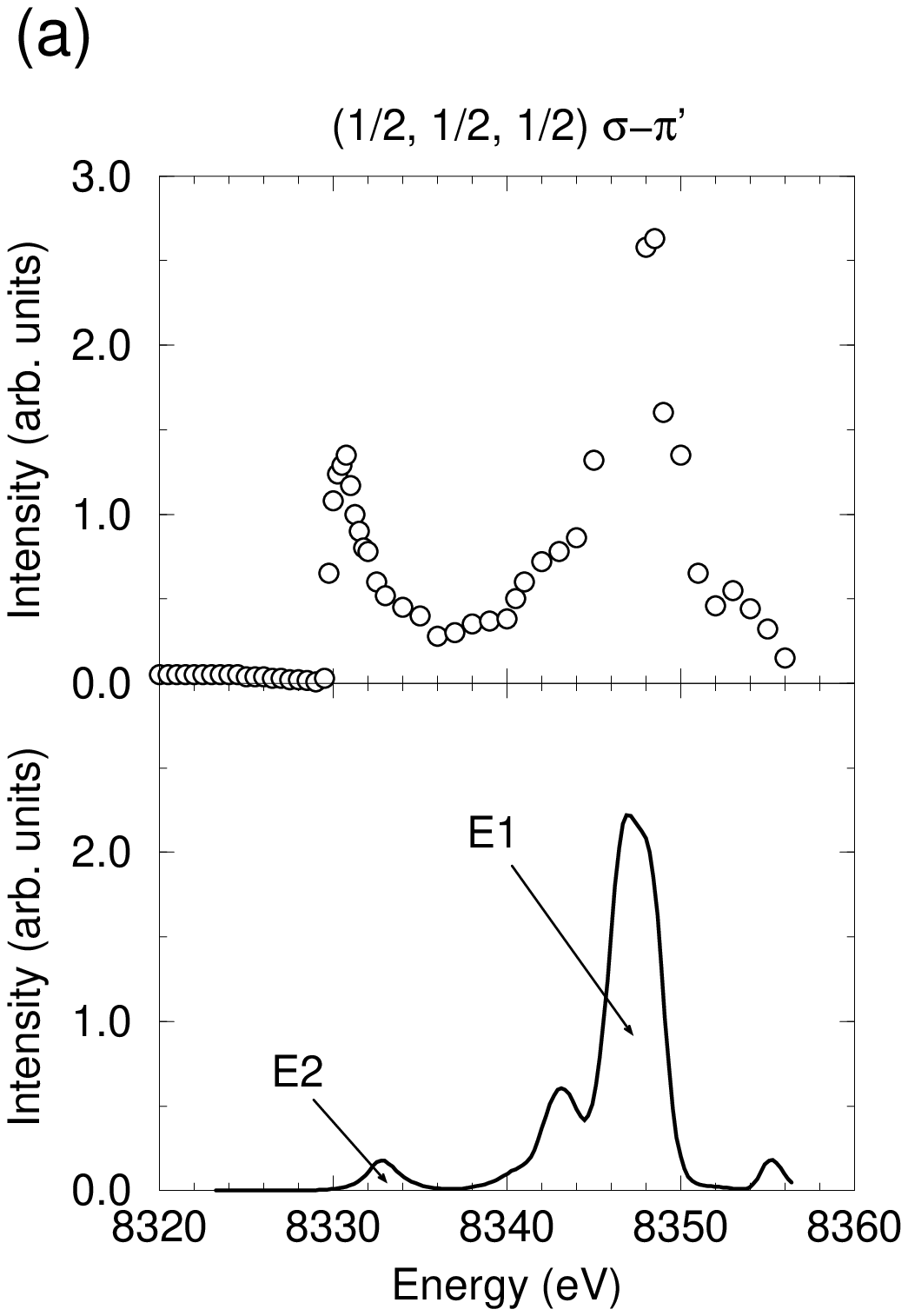}
\includegraphics[width=.72\linewidth]{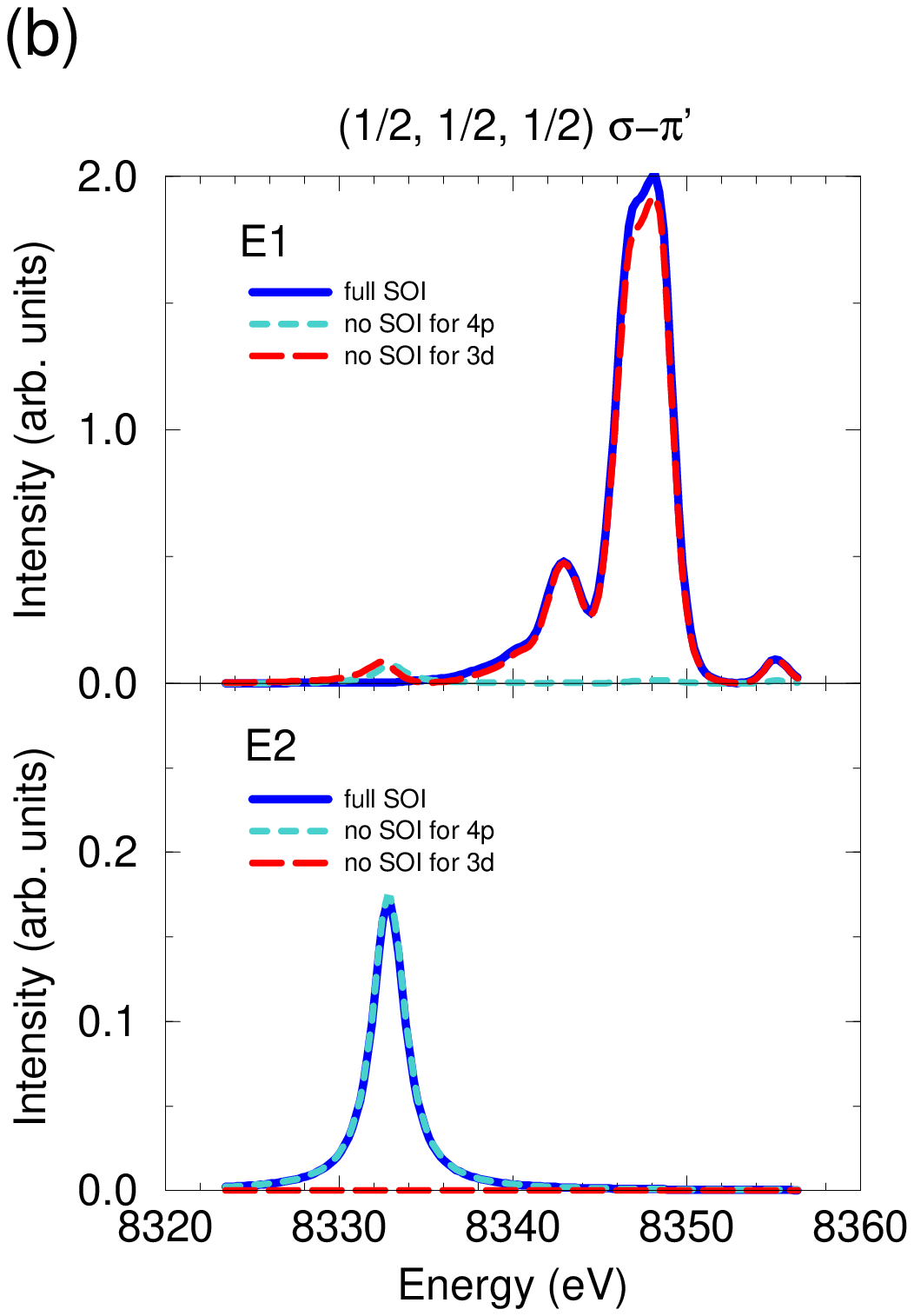}

\caption{ (color online) 
 (a) Magnetic scattering spectra in the $\sigma\to\pi'$ channel for
 ${\bf G}=(\frac{1}{2}\frac{1}{2}\frac{1}{2})$, as a function of photon
 energy.  The upper panel shows the experimental spectrum at $\psi=0^\circ$
 with three domains mixed and the absorption correction is performed.
 \cite{Neubeck3} The lower panel shows the calculated spectrum at 
 $\psi=270^\circ$ for the domain of ${\bf S}_1$. Resonant E1, E2, and 
 nonresonant contributions are included in this calculation.
 (b) Magnetic scattering spectra by turning the SOI off separately 
 on the $3d$ states and the $4p$ states, respectively.  
 Nonresonant contribution is not included in these spectra.
 \label{fig.spec1}}
\end{figure}
}

\providecommand{\FIGSEVEN}{
\begin{figure}[ht]
\includegraphics[width=.76\linewidth]{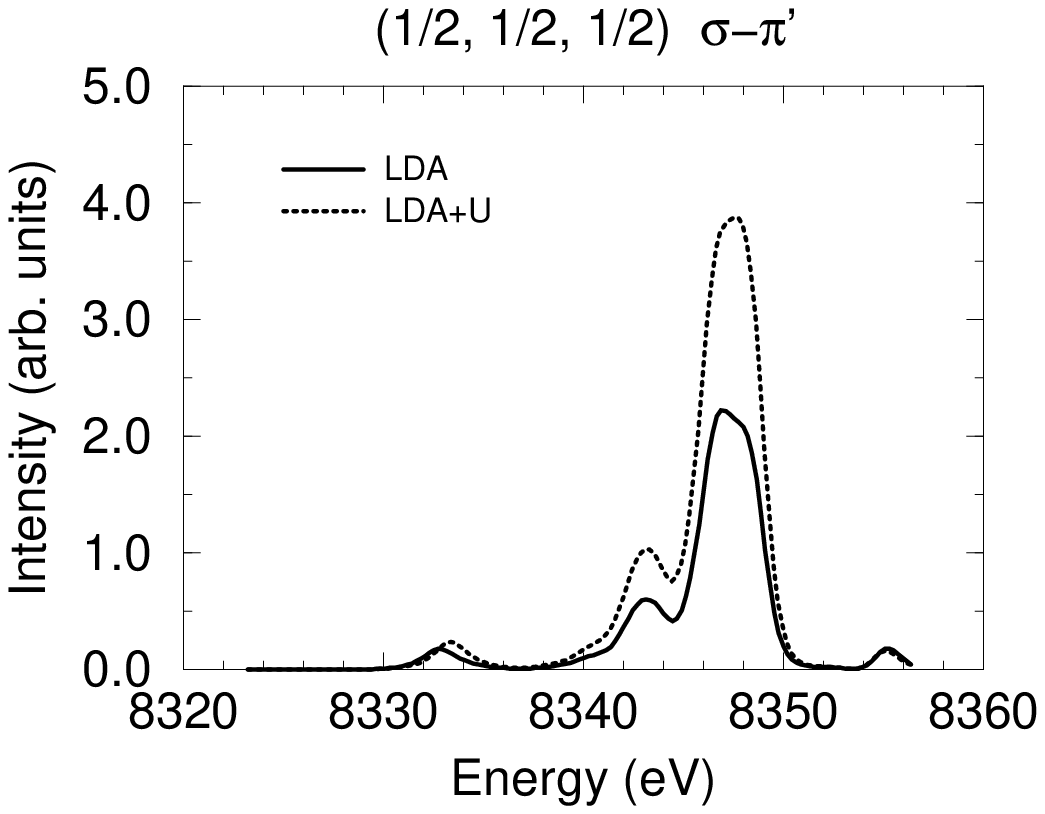}
\caption{Comparison between the RXS spectra calculated by the LDA (solid
 line) and the LDA+$U$ with $U_{\rm eff}=2$ eV (dashed line).
  \label{fig.spec1_u}}
\end{figure}
}

\providecommand{\FIGEIGHT}{
\begin{figure}[ht]
\includegraphics[width=.74\linewidth]{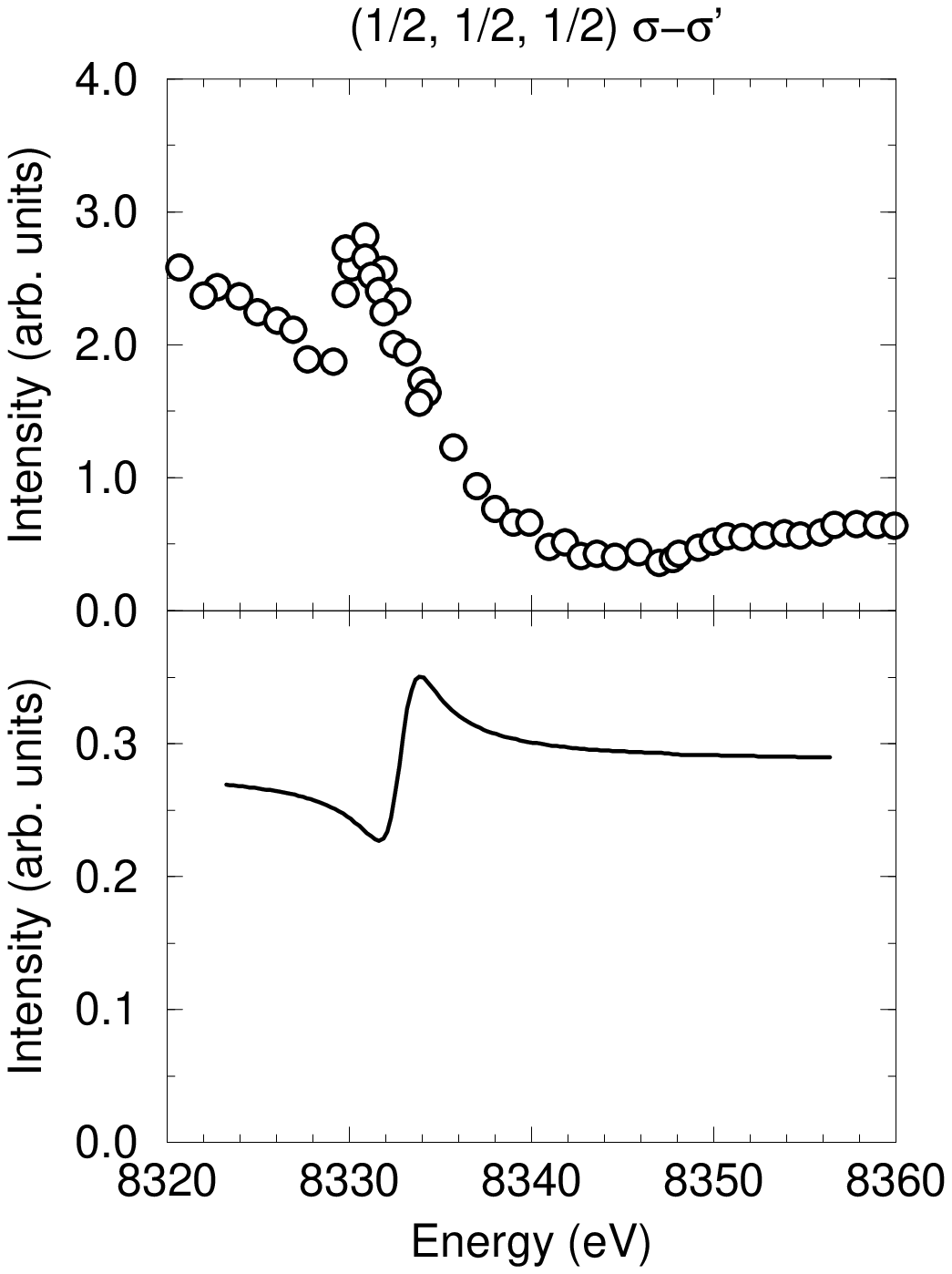}
\caption{ Magnetic scattering spectra in the $\sigma\to\sigma'$ channel for
 ${\bf G}=(\frac{1}{2}\frac{1}{2}\frac{1}{2})$, as a function of photon
 energy.  The upper panel shows the experimental spectra at $\psi=0$ without
 absorption correction.\cite{Neubeck2} The lower panel shows the calculated
 results at $\psi=0$.  \label{fig.spec2}}
\end{figure}
}

\providecommand{\FIGNINE}{
\begin{figure}[ht]
 \includegraphics[width=.7\linewidth]{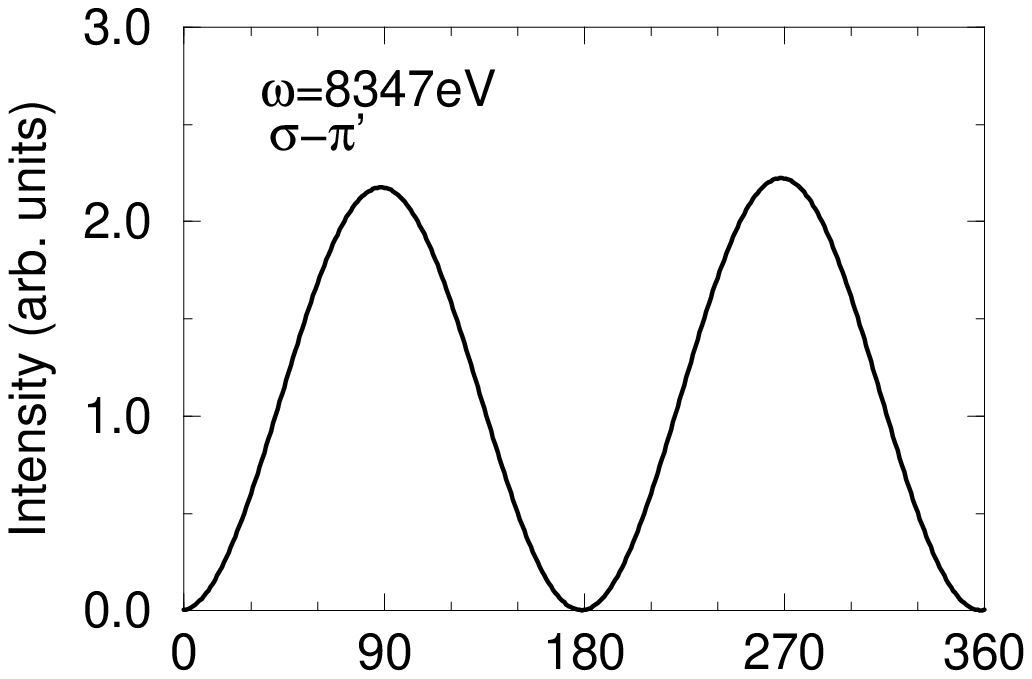}
 \includegraphics[width=.7\linewidth]{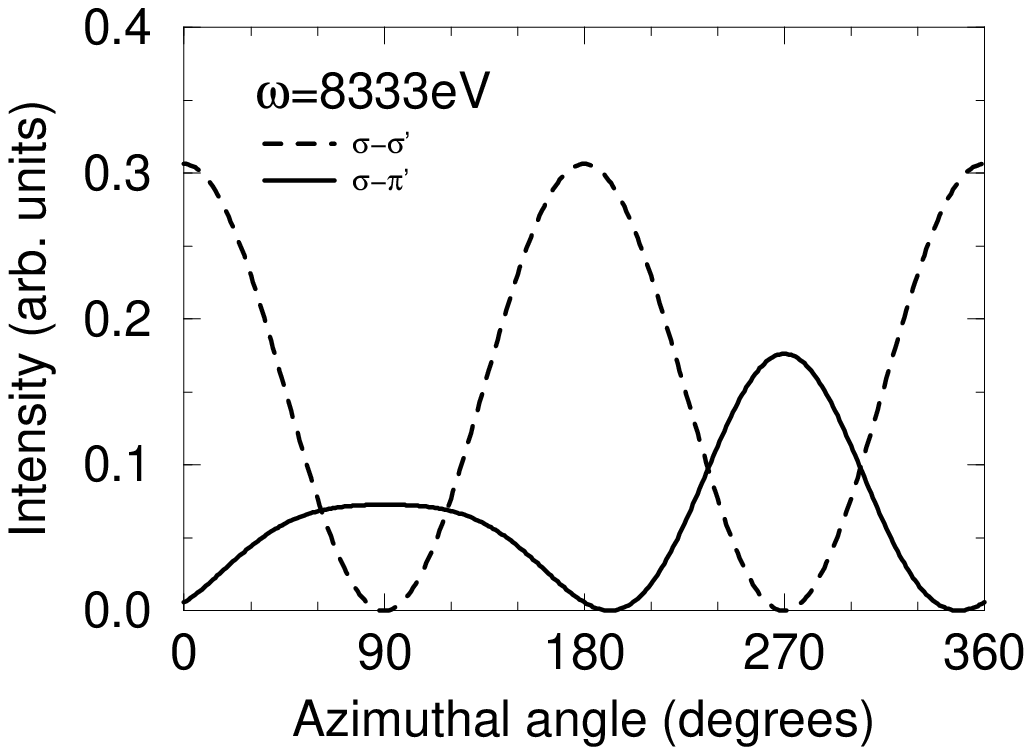}
\caption{ Azimuthal-angle dependence of the magnetic scattering intensity
 for ${\bf G}=(\frac{1}{2}\frac{1}{2}\frac{1}{2})$.  The upper panel is for
 the pre-edge peak at 8333 eV (non-resonant contribution is
 eliminated).  The lower panel is for the main peak at 8347 eV. 
 \label{fig.azim}}
\end{figure}
}

\documentclass[twocolumn,showpacs,preprintnumbers,amsmath,amssymb,prb]{revtex4}

\usepackage{graphicx}
\usepackage{dcolumn}
\usepackage{bm}

\newcommand{\beps} {\mbox{${\mbox{\boldmath $\hat{\epsilon}$}}$}}

\begin{document}



\title{An \textit{ab-initio} calculation of 
       magnetic resonant x-ray scattering spectra in NiO}


\author{Manabu Usuda}
\author{$^*$Manabu Takahashi}
\author{Jun-ichi Igarashi}
   \affiliation{%
   Synchrotron Radiation Research Center, 
   Japan Atomic Energy Research Institute, 
   Hyogo 679-5148, Japan \\
   $^*$Faculty of Engineering, Gunma University, Kiryu, 
   Gunma 376-8515, Japan}%


\date{\today} 


\begin{abstract}
We investigate the magnetic resonant x-ray scattering spectra around
 the $K$ edge of Ni in antiferromagnetic NiO using an \textit{ab-initio}
 band-structure calculation based on the density-functional theory.  By
 taking account of orbital polarization through the spin-orbit interaction,
 we reproduce well the spectra obtained experimentally, thus demonstrating 
 the usefulness of the \textit{ab-initio} calculation.  It is shown that 
 the main-edge peak, which mainly comes from the dipolar ($1s \to 4p$) 
 transition, is a direct reflection of the orbital polarization of the
 $4p$ states. It is clarified that the $4p$ orbital polarization is mainly
 induced from the spin polarization on the $4p$ states by the spin-orbit 
 interaction. The $3d$ orbital polarization at neighboring Ni sites gives 
 rise to only a minor contribution to the $4p$ orbital polarization through 
 a $p$-$d$ mixing. It is also shown that the pre-edge peak, which mainly 
 originates from the quadrupolar ($1s \to 3d$) transition, is a direct 
 reflection of the orbital polarization of the unoccupied $3d$ states. 
 It shows a Fano-type antiresonant dip due to interference with the 
 nonresonant contribution, in agreement with the experimental result.
\end{abstract}


\maketitle


\section{\label{sec:intro}Introduction}

Resonant x-ray scattering (RXS) is a powerful tool for investigating the
 properties of magnetic and orbital orders.  Resonant enhancements on
 antiferromagnetic (AF) superlattice spots have been observed in several
 transition-metal compounds such as RbMnF$_3$ (Ref.~\onlinecite{Stunault}),
 KCuF$_3$ (Ref.~\onlinecite{Caciuffo}), CoO (Ref.~\onlinecite{Neubeck1}) and
 NiO (Refs.~\onlinecite{Hill,Neubeck2}).  RXS is described as a second-order
 process wherein a photon is virtually absorbed upon exciting a core electron
 to unoccupied states, and then emitted upon recombining the excited electron
 with the core hole.  For the $3d$ transition-metal $K$-edge, the
 intermediate states involve unoccupied $4p$ states in the dipolar (E1)
 transition and unoccupied $3d$ states in the quadrupolar (E2) transition.
 The RXS intensities due to the E1 and E2 transitions constitute main-edge
 and pre-edge structures, respectively.  The main-edge intensity is 
 sensitive to the electronic structure of neighboring sites since 
 the unoccupied $4p$ states are highly extended in space.  
 This characteristic is now well recognized in RXS on orbital superlattice
 spots in the $K$ edge of transition-metal compounds. At an early stage,
 a mechanism was proposed, according to which the intensity is related to
 the $4p$ states modified by the intra-atomic Coulomb interaction with 
 the $3d$ electrons constituting the orbital order\cite{Murakami,Ishihara},
 but extensive studies based on band-structure calculations 
 \cite{Elfimov, Benfatto, Takahashi1} revealed that the intensity is 
 related to the $4p$ states modified by the lattice distortion.

In our previous work,\cite{Igarashi1} we studied the mechanism of 
 magnetic RXS in the antiferromagnetic phase of NiO, and analyzed the
 experiment performed by Neubeck \textit{et al.}\cite{Neubeck2} 
 To deal with the extended nature of the $4p$ states, we used a tight-binding 
 model.  Since orbitals must couple to the magnetic order to generate 
 the intensity of the magnetic RXS, it is essential to take account of 
 the spin-orbit interaction (SOI). The spin polarization alone in the $4p$
 states cannot give rise to the intensity, because, if we add the amplitudes 
 due to up-spin electrons to that due to down-spin electrons, the total 
 becomes the same at each site, leading to a cancellation of amplitudes 
 between different magnetic sublattices. Through the model calculation, 
 we clarified that the main-edge intensity is generated from the orbital 
 polarization in the $4p$ states. Generally speaking, such polarization is 
 brought about by the $4p$ SOI, by the intra-atomic Coulomb interaction 
 between the $4p$ and $3d$ states, and by the mixing of the $4p$ states with 
 the $3d$ states of neighboring Ni atoms. Although we obtained the magnetic 
 RXS spectra that were in qualitative agreement with the experiment, it is 
 still unclear how the $4p$ states are orbitally polarized. In this paper, 
 to eliminate the uncertainty of the model calculation, and to clarify 
 the mechanism of the magnetic RXS, we carry out an \textit{ab-initio} 
 calculation of the RXS spectra using a band-structure calculation based 
 on the density-functional theory. We take account of the SOI to deal with 
 the orbital polarization.  We neglect the attractive interaction between 
 the photoexcited electron and the $1s$ core hole. This effect on the $4p$ 
 states is known to be small, from the analysis of the magnetic circular 
 dichroism (MCD) \cite{Schutz,Ebert,Igarashi2} and of the RXS on orbital 
 superlattice spots at the $K$ edge of transition metals.
 \cite{Benfatto,Takahashi1,Takahashi2} 

In this paper, we show that the \textit{ab-initio} calculation reproduces 
 well the experimental spectra.  Moreover, we calculate the spectra by 
 turning the SOI on and off selectively in several states in order to clarify
 how the $4p$ orbital polarization is brought about. The result is that 
 the main-edge intensity is slightly reduced when the SOI on the $3d$ states
 is turned off, while it is largely diminished when the SOI on the $4p$ states
 is turned off.  This indicates that the $3d$ orbital polarization has little 
 influence on the $4p$ orbital polarization.  The same conclusion has been 
 drawn on the basis of an \textit{ab-initio} calculation of the magnetic RXS 
 spectra for KCuF$_3$ (Ref.~\onlinecite{Takahashi2}) and a recent analysis of
 MCD in Mn$_3$GaC (Ref.~\onlinecite{Takahashi3}).

We also calculate the contribution of the E2 and nonresonant processes.
 We obtain a single peak in the pre-edge region.
 It is shown to originate from the orbital polarization in the $3d$ states.
 Although the $3d$ orbital polarization is much larger than the $4p$ orbital
 polarization, the obtained pre-edge peak is rather small compared to 
 the main-edge peak.  
 The peak position is at a higher energy than the experimental one.
 We can ascribe 
 this discrepancy to the neglect of the interaction between the $1s$ core hole 
 and the $3d$ electrons in the present calculation.  We also reproduce well 
 the Fano-type antiresonant dip due to the interference between the E2 and 
 the nonresonant contributions.

This paper is organized as follows. In Sec.~\ref{sec.band}, we show the 
 results of band-structure calculations and discuss the ground state in the
 antiferromagnetic phase of NiO. In Section \ref{sec.method}, we briefly 
 describe the formulae used in calculating the magnetic RXS spectra.  In
 Sec.~\ref{sec.results}, we present the calculated magnetic RXS spectra and
 compare them with experimental data. Section \ref{sec.conclusion} is devoted
 to concluding remarks.


\section{Electronic structure of Nickel Oxide} 
\label{sec.band}

NiO has the NaCl-type structure with the lattice constant of $a=4.177$
 \AA \ (Ref.~\onlinecite{Neubeck3}).  
 Nickel atoms form an fcc lattice, as illustrated in
 Fig.~\ref{fig.cryst}. Type-II AF order develops below $T_{\rm N} = 523$ K.
  The order parameter is characterized by a wave vector ${\bf Q}$, which is
 directed to one of four body-diagonals in the fcc lattice.  We selectively
 assume that ${\bf Q} = (\frac{1}{2} \frac{1}{2} \frac{1}{2} )$ in the present
 study. The magnetic moment lies in the plane perpendicular to the AF
 modulation direction; three directions, ${\bf S}_1 = [-1,-1,2]$, ${\bf S}_2
 = [2,-1,-1]$, or ${\bf S}_3 = [-1,2,-1]$ can be taken for ${\bf Q} =
 (\frac{1}{2} \frac{1}{2} \frac{1}{2})$.  Note that the experiment of the
 magnetic RXS in NiO has been performed under the condition that ${\bf Q}$ is
 specified but domains are formed with respect to the spin
 direction~\cite{Neubeck2,Neubeck3}.

The electronic band structure of NiO is calculated using the full-potential
 linearized augmented-plane-wave (FLAPW) method \cite{Takeda79, Jansen84} in
 the local-density approximation (LDA).  The local exchange-correlation
 functional of Vosko, Wilk and Nusair~\cite{VWN80} is employed.
The experimental lattice constant is used and the muffin-tin (MT) radii are
 set to be 2.2 a.u. for nickel and 1.6 a.u. for oxygen.  We also perform
 LDA$+U$ calculation ~\cite{Anisimov, Solovyev} with the effective Coulomb
 parameter $U_{\rm eff}=U-J$ for the Ni $3d$ states, in order to investigate
 the effect of the band gap and the magnetic moment on the RXS spectra.

\FIGONE

Table~\ref{tab.1} shows the calculated values of the band gap and the
 magnetic moments in comparison with the experiments.
 LDA calculation considerably underestimates the magnetic moment
 and band gap magnitudes. Introducing $U_{\rm eff}$ enhances them but the
 obtained magnitudes are still smaller than those in the experiments.

\TABLEI

Figure \ref{fig.dos} shows the density of states (DOS) projected on the $p$
 and $d$ symmetric states inside the MT sphere at a Ni site, calculated in
 the LDA.  The origin of the energy is set at the top of the valence band.
  The right part of the figure is shown on a 20-times magnified scale.  The
 $p$ and $d$ symmetric states are sometimes referred to as the $4p$ and $3d$
 states, respectively, because the wave functions are similar to atomic
 $4p$ and $3d$ wave functions inside the MT sphere.  The magnitude of the
 $4p$ DOS is much smaller than that of the $3d$ DOS, because most of the $4p$
 density is distributed over the interstitial region or on neighboring sites.
 Introducing large $U_{\rm eff}$, such as $U_{\rm eff}=5$ eV, in the LDA$+U$
 calculation considerably changes the $3d$ DOS (not shown), while the $4p$
 DOS remains almost the same. Also, exchange splitting in the $4p$ DOS is 
 negligibly small in contrast with the large exchange splitting in the $3d$ 
 states.  Both observations suggest that the effect of the $3d$ states on 
 the $4p$ states is very small.

\FIGTWO


\section{Formulation for magnetic RXS}
\label{sec.method}

We consider the situation that a photon with the energy $\hbar\omega$, the
 momentum ${\bf k}_i$, and the polarization $\mu$ ($=\sigma$ or $\pi$) with
 the polarization vector $\beps$, is scattered into the state with the same
 energy $\hbar\omega$ (elastic scattering), the momentum ${\bf k}_f$, and
 the polarization $\mu'$ ($=\sigma'$ or $\pi'$) with the polarization vector
 $\beps'$.  The scattering geometry is depicted in Fig.~\ref{fig.geom}.  The
 magnetic scattering amplitude is derived from Fermi's golden rule
 to the second order~\cite{Arola,Rennert1,Rennert2,deBergevin1,Blume1,Blume2}, 
 in which the amplitude consists of resonant and nonresonant scatterings.
 We consider the resonant E1 and E2 transitions and
 the nonresonant contribution.
 The cross section for the elastic scattering at a magnetic superlattice
 spot is expressed as
\begin{equation}
 \frac{d\sigma}{d\Omega} 
 \propto
  \left| T_{\mu\to\mu'}({\bf G},\omega) 
       + J_{\mu\to\mu'}({\bf G},\omega) 
       + L_{\mu\to\mu'}({\bf G},\omega) \right|^2 ,
\end{equation}
where ${\bf G}={\bf k}_f-{\bf k}_i$ is a scattering vector.

The nonresonant term $T_{\mu\to\mu'}({\bf G},\omega)$ is given by
\begin{equation}
  T_{\mu\to\mu'}({\bf G},\omega) =
    -\frac{i\hbar\omega}{mc^2}\left(\frac{1}{2}{\bf L}({\bf G})
      \cdot {\bf A}''+{\bf S}({\bf G})\cdot{\bf B}\right),
\label{eq.nonreso}
\end{equation}
where $m$ is the electron mass and $c$ is the light velocity.
The vectors ${\bf A}''$ and ${\bf B}$ are given by 
\begin{eqnarray}
 {\bf A}''&=& {\bf A}'-({\bf A}'\cdot{\bf\hat G}){\bf\hat G},
  \quad {\bf A}'=-4\sin^2\theta(\beps'\times \beps),\\
 {\bf B} &=& 
 \beps'\times \beps
          + ({\bf\hat k}_f \times \beps')
            ({\bf\hat k}_f \cdot  \beps)
          - ({\bf\hat k}_i \times \beps)
            ({\bf\hat k}_i \cdot  \beps') 
         \nonumber \\
      & & - ({\bf\hat k}_f\times \beps')
     \times ({\bf\hat k}_i\times \beps),
\end{eqnarray}
where ${\bf\hat k}_i={\bf k}_i/|{\bf k}_i|$, 
     ${\bf\hat k}_f={\bf k}_f/|{\bf k}_f|$,
and  ${\bf\hat G}  ={\bf G}  /|{\bf G}|$.
The quantities ${\bf S}({\bf G})$ and ${\bf L}({\bf G})$ 
in Eq.(\ref{eq.nonreso}) are the Fourier transforms
of the spin and orbital densities, respectively~\cite{Blume2}.
Their approximate forms for calculation are given in the next section.

\FIGTHREE

The resonant contribution from the E1 transition is given by
\begin{equation}
  J_{\mu\to\mu'}({\bf G},\omega)
   = \sum_{\alpha\alpha'}({P'}^{\mu'})_{\alpha}
    M_{\alpha\alpha'}({\bf G},\omega)P^{\mu}_{\alpha'},
\label{eq.dipole}
\end{equation}
with
\begin{eqnarray}
\lefteqn{
 M_{\alpha\alpha'}({\bf G},\omega) 
}\hspace{0.1cm} \nonumber \\
 &=& 
  \frac{1}{\sqrt{N}}\sum_{n,\Lambda} e^{-i{\bf G}\cdot{\bf R}_n}
  \frac{m\omega_{\Lambda g}^2\langle g|x_\alpha(n)|\Lambda\rangle
  \langle \Lambda|x_{\alpha'}(n)|g\rangle}
       {\hbar\omega-\omega_{\Lambda g}+i\Gamma} , \nonumber \\
\end{eqnarray}
where $\omega_{\Lambda g} = (E_{\Lambda} - E_g)/\hbar$.  This represents a
 second-order process wherein a photon is virtually absorbed upon exciting the
 $1s$ electron to the $4p$ states and then emitted upon recombining the
 excited electron with the $1s$ core hole.  The geometrical factors $P^\mu$,
 ${P'}^{\mu'}$ are explicitly written in the Appendix of
 Ref.~\onlinecite{Igarashi1}.  $E_g$ represents the ground-state energy.
  The intermediate state $|\Lambda\rangle$ consists of an excited electron
 on the $4p$ states and a hole on the $1s$ state with energy $E_{\Lambda}$.
 $N$ represents the number of Ni sites, and $n$ is the index for the Ni sites.
 The dipole operators $x_\alpha(n)$ at site $n$ are defined as
 $x_1(n) = x$, $x_2(n) = y$, $x_3(n) = z$ in the coordinate frame fixed to 
 the crystal axes with the origin located at the center of site $n$. 
 Note that the amplitudes on magnetic spot ${\bf G} = (\frac{h}{2}
 \frac{k}{2} \frac{\ell}{2})$ are nearly canceled out between different 
 magnetic sublattices.  Only small values remain in an antisymmetric form,
\begin{equation}
 M({\bf G},\omega)
   = \left( \begin{array}{rrr}
            0 & a & c \\
           -a & 0 & b \\
           -c &-b & 0
            \end{array} \right) ,
\end{equation}
with $a$, $b$, and $c$ being some complex numbers.

The resonant contribution from the E2 transition is given by
\begin{equation}
  L_{\mu\to\mu'}({\bf G},\omega)
   = \sum_{\gamma\gamma'}({Q'}^{\mu'})_{\gamma}
    N_{\gamma\gamma'}({\bf G},\omega)Q^{\mu}_{\gamma'},
\label{eq.quadrupole}
\end{equation}
with
\begin{eqnarray}
\lefteqn{
 N_{\gamma\gamma'}({\bf G},\omega) = 
  \frac{1}{\sqrt{N}}\sum_{n,\Lambda'} e^{-i{\bf G}\cdot{\bf R}_n} 
}\hspace{0.1cm}\nonumber \\
 & & \times \frac{1}{12}\left(\frac{\omega}{c}\right)^2
  \frac{m\omega_{\Lambda'g}^2\langle g|z_\gamma(n)|\Lambda'\rangle
 \langle\Lambda'
       |z_{\gamma'}(n)|g\rangle}
       {\hbar\omega-\omega_{\Lambda'g}+i\Gamma}, \nonumber \\
\end{eqnarray}
where $\omega_{\Lambda' g} = (E_{\Lambda'} - E_g)/\hbar$.
The geometrical factors $Q^{\mu}$, ${Q'}^{\mu'}$ are given in the Appendix
 of Ref.~\onlinecite{Igarashi1}.  The intermediate states $|\Lambda'\rangle$
 consist of an excited electron on the $3d$ states and a hole on the $1s$
 state with energy $E_{\Lambda'}$.
 The quadrupole operators $z_\mu(n)$ are defined as
 $z_1(n) = (\sqrt{3}/2) (x^2 - y^2)$, 
 $z_2(n) = (1/2) (3z^2 - r^2)$, 
 $z_3(n) = \sqrt{3} yz$, 
 $z_4(n) = \sqrt{3} zx$,
 $z_5(n) = \sqrt{3} xy$.
  Similar to the E1 transition, the amplitudes on magnetic spot ${\bf G}
 = (\frac{h}{2} \frac{k}{2} \frac{\ell}{2})$ are nearly canceled out
 between different magnetic sublattices, resulting in an antisymmetric form,
\begin{equation}
 N({\bf G},\omega)
   = \left( \begin{array}{rrrrr}
             0 & d & e & f & g \\
            -d & 0 & u & p & q \\
            -e &-u & 0 & r & s \\
            -f &-p &-r & 0 & t \\ 
            -g &-q &-s &-t & 0 
          \end{array} \right) , 
\end{equation} 
with $d\sim u$ being some complex numbers.

\section{MAGNETIC RXS SPECTRA}
\label{sec.results}

\subsection{Absorption coefficient}

The absorption coefficient within the E1 transition is given by
\begin{equation}
 A(\omega) \propto \sum_{\Lambda,\alpha}
    |\langle\Lambda| x_\alpha(j) |g\rangle|^2
    \frac{\Gamma}{(\hbar\omega-E_\Lambda)^2+\Gamma^2}.
\end{equation}
We neglect the core-hole potential working on the $4p$ states in the final
 state of the absorption process, and replace $|\Lambda\rangle$ by the
 unoccupied $4p$ states given by the band-structure calculation.  The
 transition matrix elements are evaluated using the LDA wave-functions.  The
 core-hole energy is adjusted such that the calculated peak-position
 coincides with the experimental one ($\hbar\omega=8347$ eV for the peak
 \cite{Neubeck1}).  We set the core-hole lifetime width $\Gamma$ as 1 eV.
  Figure \ref{fig.absorp} shows the calculated results and those of
 the $K$-edge fluorescence experiment.\cite{Neubeck2} The experimental 
 spectral shape is reproduced well, suggesting that the core-hole potential
 plays only a minor role in the spectral shape. As already mentioned in 
 Sec.~I, this observation is consistent with the recent analysis of the 
 $K$-edge MCD spectra of Mn$_3$GaC (Ref.~\onlinecite{Takahashi3}) as well as 
 of $K$-edge RXS on orbital superlattice spots in transition-metal
 compounds.\cite{Benfatto,Takahashi1}

\FIGFOUR


\subsection{Magnetic RXS spectra}
We evaluate the nonresonant term using the ``spherical approximation''
and ``dipolar approximation'' for the Fourier transforms ${\bf S}(\bf G)$ 
and ${\bf L}({\bf G})$, respectively \cite{Coppens}.
${\bf S}(\bf G)$ is expressed as
\begin{equation}
 {\bf S}({\bf G}) \approx 
 \frac{1}{\sqrt{N}}\sum_n e^{-i{\bf G}\cdot{\bf R}_n}
  \langle j_{0n}\rangle \langle{\bf S}_n\rangle,
\end{equation}
with
\begin{equation}
\langle j_{0n}\rangle = \int_0^{R_{\rm MT}}j_0(Gr)[\phi_{nl}(r)]^2 r^2{\rm d}r,
\end{equation}
where $j_0(x)=\sin x/x$.
$R_{\rm MT}$ is the MT radius and $\phi_{nl}(r)$ is the radial function 
with $l$ being the angular momentum on the Ni $n$ site. 
$\langle{\bf S}_n\rangle$ is the average of the spin operator 
within the MT sphere at site $n$.
The $3d$ states ($l=2$) give a dominant contribution.

${\bf L}(\bf G)$ is expressed as
\begin{equation}
 {\bf L}({\bf G}) \approx \frac{1}{\sqrt{N}}\sum_n e^{-i{\bf G}\cdot{\bf R}_n}
  \langle g_{0n}\rangle \langle{\bf L}_n\rangle,
\end{equation}
with
 \begin{equation}
\langle g_{0n}\rangle = \int_0^{R_{\rm MT}}g_0(Gr)[\phi_{nl}(r)]^2 r^2{\rm d}r,
 \end{equation}
 where $g_0(x)=\frac{2}{x^2}(1-\cos x)$.  $\langle{\bf L}_n\rangle$ is the
 average of the angular momentum with respect to the angular part of the
 wave function within the MT sphere at the Ni $n$ site. Again, the $3d$
 states give the dominant contribution.  Figure \ref{fig.form} shows
 calculated $|{\bf S} ({\bf G})| / \sqrt{N}$ and $|{\bf L} ({\bf G})| /
 \sqrt{N}$ and those in the experiment \cite{Neubeck1}.  The calculated 
 values are slightly smaller than the experimental ones.  This is
 because the LDA calculation underestimates the magnetic
 moments.  The Fourier transform ${\bf S}({\bf G})$ can be directly
 evaluated without the spherical approximation from the LDA wavefunctions.
 We compare the results with those evaluated from the spherical
 approximation.  The values in units of $\hbar$ are 
$|{\bf S}({\bf G})|/\sqrt{N}=0.57 (0.56)$, $0.32 (0.33)$, and $0.10 (0.13)$ 
for
${\bf G} = (\frac{1}{2}\frac{1}{2}\frac{1}{2})$, 
$          (\frac{3}{2}\frac{3}{2}\frac{3}{2})$, and
$          (\frac{5}{2}\frac{5}{2}\frac{5}{2})$, respectively,
 where the numbers in parentheses are values obtained by the spherical 
 approximation. The spherical approximation works rather well for small 
 values of $|{\bf G}|$. A similar tendency is expected to exist for 
 ${\bf L}({\bf G})$ calculated with the dipolar approximation, according to 
 the study for Ni$^{++}$ by Blume.\cite{Blume3}

\FIGFIVE

Once ${\bf S}({\bf G})$ and ${\bf L}({\bf G})$ are obtained, the
 nonresonant contribution is easily evaluated.  Then the resonant
 contribution is evaluated by replacing the intermediate state
 $|\Lambda\rangle$ in Eq.~(3.8) and $|\Lambda'\rangle$ in Eq.~(3.11) with
 the unoccupied $p$ and $d$ symmetric states given by the band-structure
 calculation.  We confine ourselves to the calculation on a single domain of
 ${\bf S}_1$ with the AF modulation vector ${\bf Q}$ fixed at $(\frac{1}{2}
 \frac{1}{2} \frac{1}{2})$.  Figure \ref{fig.spec1}(a) shows the spectra in 
 the $\sigma\to\pi'$ channel for ${\bf G} = (\frac{1}{2} \frac{1}{2}
 \frac{1}{2})$ as a function of photon energy.  
 The lower panel of Fig.~\ref{fig.spec1}(a) shows the
 calculated spectra at the azimuthal angle $\psi = 270^\circ$, where $\psi$ 
 is defined such that the scattering plane at $\psi = 0$ contains the
 $[1,-1,0]$ crystal axis.  The core-hole energy has already been determined
 such that the calculated peak in the $K$-edge absorption coefficient
 coincides with the experimental one.  
 The upper panel of Fig.~\ref{fig.spec1}(a) shows the
 experimental curve at $\psi=0^\circ$, where the absorption correction is
 made.\cite{Neubeck3} The experiment is carried out on a sample in which
 three domains of ${\bf S}_1$, ${\bf S}_2$, and ${\bf S}_3$ are mixed with
 various relative volumes.

\FIGSIX

We obtain several peaks at around $\hbar\omega=8347$ eV due to the E1 
 transition. The calculated curve agrees well with the experimental one.  
 Such comparison makes sense, because the spectral shape is a function of 
 photon energy and depends little on domains and azimuthal angle.
 \footnote{ The relative intensities among the nonresonant term, the E1 
 transition term, and the E2 transition term depend sensitively on the 
 domains and the azimuthal angle.} 
 The E1 intensity arises from the orbital polarization in the $4p$ states.  
 To clarify how the polarization is induced, we calculate
 the spectra by turning the SOI on and off separately on the $3d$ states and
 on the $4p$ states.  The results are shown in the upper panel of 
 Fig.~\ref{fig.spec1}(b). The former slightly suppresses the E1 intensity, 
 while the latter does so strongly.
 This indicates that, for the $4p$ orbital polarization, the contribution of 
 the $3d$ orbital polarization through the $p$-$d$ hybridization is quite small
 and the $4p$ orbital polarization is mainly induced by the $4p$ spin 
 polarization  through the SOI.
 This is consistent with the recent analysis 
 of the $K$-edge MCD in the ferromagnetic phase of Mn$_3$GaC
 (Ref.~\onlinecite{Takahashi3})
 where it was clarified that the $4p$ orbital polarization near the Fermi 
 level is mainly
 induced by coupling to the $3d$ orbital polarization, while that away
 from the Fermi level is induced by the $4p$ spin polarization through the
 SOI.

In addition to the main-edge peak, we obtain another peak at around 
 $\hbar\omega = 8333$ eV. This pre-edge intensity is attributed to the
 E2 transition, that is, the transition to unoccupied $3d$ states in the 
 intermediate state.  
 As shown in the upper panel of Fig.~\ref{fig.spec1}(b),
 partially turning the SOI off generates small E1 intensities around
 $\hbar\omega=$ 8333 eV, but the net contribution from the E1 transition 
 (the solid line in the upper panel of Fig.~\ref{fig.spec1}(b))
 becomes negligible due to a cancellation.
 As shown in the lower panel of Fig.~\ref{fig.spec1}(b), 
 the E2 intensity is strongly suppressed by turning the SOI off on the $3d$ 
 states, indicating that it originates from the $3d$ orbital polarization.  
 The magnitude of the pre-edge peak is rather small compared to that of 
 the main-edge peak, though the $3d$ orbital polarization is much larger than
 the $4p$ orbital polarization.
 The experimental pre-edge peak is located at 
 $\hbar\omega = 8330$ eV, which is 3 eV lower than the calculated position. 
 We can attribute this difference to the effect of the interaction between the 
 core hole and the $3d$ states.  The attractive interaction is expected to be 
 stronger on the $3d$ states than on the $4p$ states, because the former 
 states are more localized than the latter ones.  The nonresonant contribution 
 is  extremely small at this azimuthal angle.

In Fig.~\ref{fig.spec1_u}, we show the RXS spectra calculated with the
 LDA and the LDA+$U$.  The intensity is enhanced due to the local Coulomb
 interaction $U_{\rm eff}$.  This is consistent with the enhancement of the
 orbital moment (see Table I).  The pre-edge peak is shifted to higher
 energy on introducing $U_{\rm eff}$, which correlates with the opening
 larger energy gap.  Nevertheless, the effect of $U_{\rm eff}$ on the
 whole spectral shape is small.

\FIGSEVEN

Figure \ref{fig.spec2} shows the spectra in the $\sigma\to\sigma'$ channel
 for ${\bf G}=(\frac{1}{2}\frac{1}{2}\frac{1}{2})$.  Both the calculated
 and experimental curves are for $\psi=0$.  
 The absorption correction is not made on the experimental curve, 
 so the nonresonant intensities below and above the resonant peak are not
 the same in the experimental curve~\cite{Neubeck2}.
 Since the E1 transition is forbidden for all $\psi$ in the $\sigma\to\sigma'$ 
 channel, no main-edge peak appears.  The pre-edge peak is located at an 
 $\sim 3$ eV higher position, due to the same reason as in the case of the 
 $\sigma\to\pi'$ channel. We obtain a Fano-type antiresonant dip on 
 the low-energy side of the peak, which arises from an interference with 
 the nonresonant contribution, in agreement with the experiment.
\FIGEIGHT

Finally we calculate the azimuthal-angle dependence of the intensity.
 Although the calculation is performed for a single domain and it cannot be
 directly compared with the experimental results, it may serve as a guide 
 for analyzing mixed domains.  Figure \ref{fig.azim} shows 
 the result for ${\bf G} = (\frac{1}{2} \frac{1}{2} \frac{1}{2})$.  
 The upper (lower) panel shows the intensity of 
 the main-edge (pre-edge) peak at $\hbar\omega = 8347$ eV (8333 eV).  
 The dependence of the main-edge peak is controlled by the geometrical factor
 of the E1 transition. As already mentioned, the E1 transition is completely 
 forbidden in the $\sigma \to \sigma'$ channel.  

\FIGNINE

\section{CONCLUDING REMARKS}
\label{sec.conclusion}

We presented a theoretical study of the magnetic RXS around the $K$
 edge of Ni in antiferromagnetic NiO using the FLAPW band-structure
 calculation in the density-functional theory.  The calculation reproduced
 the experimental spectra, thus demonstrating the usefulness of the
 \textit{ab-initio} calculation.  The main-edge peak originates from 
 the orbital polarization in the $4p$ states through the E1 transition. 
 We found that the $4p$ orbital polarization is mainly generated from the 
 $4p$ spin polarization through the SOI and the effect of coupling 
 to the $3d$ orbital polarization at neighboring Ni sites is found
 to be quite small. The $4p$ orbital polarization is closely related to
 the $K$-edge MCD spectra. It is known from the analysis of the MCD spectra
 in Mn$_3$GaC (Ref.~\onlinecite{Takahashi3}), and in Fe, Co and Ni
 (Ref.~\onlinecite{Schutz})
 that the MCD signals near the Fermi level are generated 
 mainly by the $4p$ orbital polarization coupled to the $3d$ orbital 
 polarization at neighboring sites, while the MCD signals well above 
 the Fermi level are generated from the $4p$ orbital polarization 
 coupled to the $4p$ spin polarization through the SOI. 
 The present finding for the main-edge peak is corresponding to
 the latter.
 
We also obtained the pre-edge intensity through the E2 transition. 
 The contribution of the E1 transition to the pre-edge intensity is 
 found to be negligible. This shows a contrast with the RXS spectra 
 as to the orbital-ordering spots in LaMnO$_3$. The E1 transition 
 generates pre-edge intensity through mixing $4p$ states with 
 neighboring $3d$ states. In LaMnO$_3$ the pre-edge intensity of 
 the E1 transition is much smaller than the main-edge intensity but 
 still larger than the pre-edge intensity of the E2 transition. 
 In the present case, the main-edge intensity on the magnetic spot 
 is much smaller than that on the orbital-ordering spot in LaMnO$_3$, 
 and so becomes the pre-edge intensity of the E1 transition. 
 Thereby the intensity of the E2 transition becomes relatively large. 
 
The calculated pre-edge peak position relative to the main-edge peak is
 $\sim 3$ eV higher than the experimental one.  This
 suggests that the core-hole potential is more effective on the $3d$ states.
 The inclusion of such an effect by extending the calculation scheme is left 
 for future studies.

\begin{acknowledgments}
We have used the FLAPW code developed by Professor N. Hamada. We thank him for
 allowing us to use his code.  We would also like to thank Dr. W. Neubeck 
 for sending us his thesis and for valuable discussion. This work was 
 partially supported by a Grant-in-Aid for Scientific Research from 
 the Ministry of Education, Science, Sports and Culture, Japan.
\end{acknowledgments}



\bibliography{NiO}

\begin{thebibliography}{33}
\expandafter\ifx\csname natexlab\endcsname\relax\def\natexlab#1{#1}\fi
\expandafter\ifx\csname bibnamefont\endcsname\relax
  \def\bibnamefont#1{#1}\fi
\expandafter\ifx\csname bibfnamefont\endcsname\relax
  \def\bibfnamefont#1{#1}\fi
\expandafter\ifx\csname citenamefont\endcsname\relax
  \def\citenamefont#1{#1}\fi
\expandafter\ifx\csname url\endcsname\relax
  \def\url#1{\texttt{#1}}\fi
\expandafter\ifx\csname urlprefix\endcsname\relax\def\urlprefix{URL }\fi
\providecommand{\bibinfo}[2]{#2}
\providecommand{\eprint}[2][]{\url{#2}}

\bibitem[{\citenamefont{Stunault et~al.}(1999)\citenamefont{Stunault,
  de~Bergevin, Wermeille, Vettier, Br{\"u}ckel, Bernhoeft, McIntyre, and
  Henry}}]{Stunault}
\bibinfo{author}{\bibfnamefont{A.}~\bibnamefont{Stunault}},
  \bibinfo{author}{\bibfnamefont{F.}~\bibnamefont{de~Bergevin}},
  \bibinfo{author}{\bibfnamefont{D.}~\bibnamefont{Wermeille}},
  \bibinfo{author}{\bibfnamefont{C.}~\bibnamefont{Vettier}},
  \bibinfo{author}{\bibfnamefont{T.}~\bibnamefont{Br{\"u}ckel}},
  \bibinfo{author}{\bibfnamefont{N.}~\bibnamefont{Bernhoeft}},
  \bibinfo{author}{\bibfnamefont{G.~J.} \bibnamefont{McIntyre}},
  \bibnamefont{and} \bibinfo{author}{\bibfnamefont{J.~Y.} \bibnamefont{Henry}},
  \bibinfo{journal}{Phys.\ Rev. B} \textbf{\bibinfo{volume}{60}},
  \bibinfo{pages}{10170} (\bibinfo{year}{1999}).

\bibitem[{\citenamefont{Caciuffo et~al.}(2002)\citenamefont{Caciuffo,
  Paolasini, Sollier, Ghigna, Pavarini, van~den Brink, and
  Altarelli}}]{Caciuffo}
\bibinfo{author}{\bibfnamefont{R.}~\bibnamefont{Caciuffo}},
  \bibinfo{author}{\bibfnamefont{L.}~\bibnamefont{Paolasini}},
  \bibinfo{author}{\bibfnamefont{A.}~\bibnamefont{Sollier}},
  \bibinfo{author}{\bibfnamefont{P.}~\bibnamefont{Ghigna}},
  \bibinfo{author}{\bibfnamefont{E.}~\bibnamefont{Pavarini}},
  \bibinfo{author}{\bibfnamefont{J.}~\bibnamefont{van~den Brink}},
  \bibnamefont{and}
  \bibinfo{author}{\bibfnamefont{M.}~\bibnamefont{Altarelli}},
  \bibinfo{journal}{Phys.\ Rev. B} \textbf{\bibinfo{volume}{65}},
  \bibinfo{pages}{174425} (\bibinfo{year}{2002}).

\bibitem[{\citenamefont{Neubeck et~al.}(1999)\citenamefont{Neubeck, Vettier,
  Lee, and de~Bergevin}}]{Neubeck1}
\bibinfo{author}{\bibfnamefont{W.}~\bibnamefont{Neubeck}},
  \bibinfo{author}{\bibfnamefont{C.}~\bibnamefont{Vettier}},
  \bibinfo{author}{\bibfnamefont{K.-B.} \bibnamefont{Lee}}, \bibnamefont{and}
  \bibinfo{author}{\bibfnamefont{F.}~\bibnamefont{de~Bergevin}},
  \bibinfo{journal}{Phys.\ Rev. B} \textbf{\bibinfo{volume}{60}},
  \bibinfo{pages}{R9912} (\bibinfo{year}{1999}).

\bibitem[{\citenamefont{Hill et~al.}(1997)\citenamefont{Hill, Kao, and
  McMorrow}}]{Hill}
\bibinfo{author}{\bibfnamefont{J.~P.} \bibnamefont{Hill}},
  \bibinfo{author}{\bibfnamefont{C.-C.} \bibnamefont{Kao}}, \bibnamefont{and}
  \bibinfo{author}{\bibfnamefont{D.~F.} \bibnamefont{McMorrow}},
  \bibinfo{journal}{Phys.\ Rev. B} \textbf{\bibinfo{volume}{55}},
  \bibinfo{pages}{R8662} (\bibinfo{year}{1997}).

\bibitem[{\citenamefont{Neubeck et~al.}(2001)\citenamefont{Neubeck, Vettier,
  de~Bergevin, Yakhou, Mannix, Bengone, Alouani, and Barbier}}]{Neubeck2}
\bibinfo{author}{\bibfnamefont{W.}~\bibnamefont{Neubeck}},
  \bibinfo{author}{\bibfnamefont{C.}~\bibnamefont{Vettier}},
  \bibinfo{author}{\bibfnamefont{F.}~\bibnamefont{de~Bergevin}},
  \bibinfo{author}{\bibfnamefont{F.}~\bibnamefont{Yakhou}},
  \bibinfo{author}{\bibfnamefont{D.}~\bibnamefont{Mannix}},
  \bibinfo{author}{\bibfnamefont{O.}~\bibnamefont{Bengone}},
  \bibinfo{author}{\bibfnamefont{M.}~\bibnamefont{Alouani}}, \bibnamefont{and}
  \bibinfo{author}{\bibfnamefont{A.}~\bibnamefont{Barbier}},
  \bibinfo{journal}{Phys.\ Rev. B} \textbf{\bibinfo{volume}{63}},
  \bibinfo{pages}{134430} (\bibinfo{year}{2001}).

\bibitem[{\citenamefont{Murakami et~al.}(1998)\citenamefont{Murakami, Hill,
  Gibbs, Blume, Koyama, Tanaka, Kawata, Arima, Tokura, Hirota
  et~al.}}]{Murakami}
\bibinfo{author}{\bibfnamefont{Y.}~\bibnamefont{Murakami}},
  \bibinfo{author}{\bibfnamefont{J.~P.} \bibnamefont{Hill}},
  \bibinfo{author}{\bibfnamefont{D.}~\bibnamefont{Gibbs}},
  \bibinfo{author}{\bibfnamefont{M.}~\bibnamefont{Blume}},
  \bibinfo{author}{\bibfnamefont{I.}~\bibnamefont{Koyama}},
  \bibinfo{author}{\bibfnamefont{M.}~\bibnamefont{Tanaka}},
  \bibinfo{author}{\bibfnamefont{H.}~\bibnamefont{Kawata}},
  \bibinfo{author}{\bibfnamefont{T.}~\bibnamefont{Arima}},
  \bibinfo{author}{\bibfnamefont{Y.}~\bibnamefont{Tokura}},
  \bibinfo{author}{\bibfnamefont{K.}~\bibnamefont{Hirota}},
  \bibnamefont{et~al.}, \bibinfo{journal}{Phys.\ Rev.\ Lett.}
  \textbf{\bibinfo{volume}{81}}, \bibinfo{pages}{582} (\bibinfo{year}{1998}).

\bibitem[{\citenamefont{Ishihara and Maekawa}(1998)}]{Ishihara}
\bibinfo{author}{\bibfnamefont{S.}~\bibnamefont{Ishihara}} \bibnamefont{and}
  \bibinfo{author}{\bibfnamefont{S.}~\bibnamefont{Maekawa}},
  \bibinfo{journal}{Phys.\ Rev.\ Lett.} \textbf{\bibinfo{volume}{80}},
  \bibinfo{pages}{3799} (\bibinfo{year}{1998}).

\bibitem[{\citenamefont{Elfimov et~al.}(1999)\citenamefont{Elfimov, Anisimov,
  and Sawatzky}}]{Elfimov}
\bibinfo{author}{\bibfnamefont{I.~S.} \bibnamefont{Elfimov}},
  \bibinfo{author}{\bibfnamefont{V.~I.} \bibnamefont{Anisimov}},
  \bibnamefont{and} \bibinfo{author}{\bibfnamefont{G.}~\bibnamefont{Sawatzky}},
  \bibinfo{journal}{Phys.\ Rev.\ Lett.} \textbf{\bibinfo{volume}{82}},
  \bibinfo{pages}{4264} (\bibinfo{year}{1999}).

\bibitem[{\citenamefont{Benfatto et~al.}(1999)\citenamefont{Benfatto, Joly, and
  Natori}}]{Benfatto}
\bibinfo{author}{\bibfnamefont{M.}~\bibnamefont{Benfatto}},
  \bibinfo{author}{\bibfnamefont{Y.}~\bibnamefont{Joly}}, \bibnamefont{and}
  \bibinfo{author}{\bibfnamefont{C.~R.} \bibnamefont{Natori}},
  \bibinfo{journal}{Phys.\ Rev.\ Lett.} \textbf{\bibinfo{volume}{83}},
  \bibinfo{pages}{636} (\bibinfo{year}{1999}).

\bibitem[{\citenamefont{Takahashi et~al.}(1999)\citenamefont{Takahashi,
  Igarashi, and Fulde}}]{Takahashi1}
\bibinfo{author}{\bibfnamefont{M.}~\bibnamefont{Takahashi}},
  \bibinfo{author}{\bibfnamefont{J.}~\bibnamefont{Igarashi}}, \bibnamefont{and}
  \bibinfo{author}{\bibfnamefont{P.}~\bibnamefont{Fulde}}, \bibinfo{journal}{J.
  Phys. Soc. Jpn.} \textbf{\bibinfo{volume}{68}}, \bibinfo{pages}{2530}
  (\bibinfo{year}{1999}).

\bibitem[{\citenamefont{Igarashi and Takahashi}(2001)}]{Igarashi1}
\bibinfo{author}{\bibfnamefont{J.}~\bibnamefont{Igarashi}} \bibnamefont{and}
  \bibinfo{author}{\bibfnamefont{M.}~\bibnamefont{Takahashi}},
  \bibinfo{journal}{Phys.\ Rev. B} \textbf{\bibinfo{volume}{63}},
  \bibinfo{pages}{184430} (\bibinfo{year}{2001}).

\bibitem[{\citenamefont{Sch{\"u}tz et~al.}(1987)\citenamefont{Sch{\"u}tz,
  Wagner, Wilhelm, Kienle, Zeller, Frahn, and Materlik}}]{Schutz}
\bibinfo{author}{\bibfnamefont{G.}~\bibnamefont{Sch{\"u}tz}},
  \bibinfo{author}{\bibfnamefont{W.}~\bibnamefont{Wagner}},
  \bibinfo{author}{\bibfnamefont{W.}~\bibnamefont{Wilhelm}},
  \bibinfo{author}{\bibfnamefont{P.}~\bibnamefont{Kienle}},
  \bibinfo{author}{\bibfnamefont{R.}~\bibnamefont{Zeller}},
  \bibinfo{author}{\bibfnamefont{R.}~\bibnamefont{Frahn}}, \bibnamefont{and}
  \bibinfo{author}{\bibfnamefont{G.}~\bibnamefont{Materlik}},
  \bibinfo{journal}{Phy.\ Rev.\ Lett.} \textbf{\bibinfo{volume}{58}},
  \bibinfo{pages}{737} (\bibinfo{year}{1987}).

\bibitem[{\citenamefont{Ebert et~al.}(1988)\citenamefont{Ebert, Strange, and
  Gyorffy}}]{Ebert}
\bibinfo{author}{\bibfnamefont{H.}~\bibnamefont{Ebert}},
  \bibinfo{author}{\bibfnamefont{P.}~\bibnamefont{Strange}}, \bibnamefont{and}
  \bibinfo{author}{\bibfnamefont{B.~L.} \bibnamefont{Gyorffy}},
  \bibinfo{journal}{J. Appl. Phys.} \textbf{\bibinfo{volume}{63}},
  \bibinfo{pages}{3055} (\bibinfo{year}{1988}).

\bibitem[{\citenamefont{Igarashi and Hirai}(1996)}]{Igarashi2}
\bibinfo{author}{\bibfnamefont{J.}~\bibnamefont{Igarashi}} \bibnamefont{and}
  \bibinfo{author}{\bibfnamefont{K.}~\bibnamefont{Hirai}},
  \bibinfo{journal}{Phys.\ Rev.\ B} \textbf{\bibinfo{volume}{50}},
  \bibinfo{pages}{17820} (\bibinfo{year}{1996}).

\bibitem[{\citenamefont{Takahashi et~al.}(2003)\citenamefont{Takahashi, Usuda,
  and Igarashi}}]{Takahashi2}
\bibinfo{author}{\bibfnamefont{M.}~\bibnamefont{Takahashi}},
  \bibinfo{author}{\bibfnamefont{M.}~\bibnamefont{Usuda}}, \bibnamefont{and}
  \bibinfo{author}{\bibfnamefont{J.}~\bibnamefont{Igarashi}},
  \bibinfo{journal}{Phys.\ Rev.\ B} \textbf{\bibinfo{volume}{67}},
  \bibinfo{pages}{064425} (\bibinfo{year}{2003}).

\bibitem[{\citenamefont{Takahashi and Igarashi}(2003)}]{Takahashi3}
\bibinfo{author}{\bibfnamefont{M.}~\bibnamefont{Takahashi}} \bibnamefont{and}
  \bibinfo{author}{\bibfnamefont{J.}~\bibnamefont{Igarashi}},
  \bibinfo{journal}{Phys.\ Rev.\ B} \textbf{\bibinfo{volume}{67}},
  \bibinfo{pages}{245104} (\bibinfo{year}{2003}).

\bibitem[{\citenamefont{Neubeck}(2000)}]{Neubeck3}
\bibinfo{author}{\bibfnamefont{W.}~\bibnamefont{Neubeck}}, Ph.D. thesis,
  \bibinfo{school}{Joseph Fourier University, Grenoble I}
  (\bibinfo{year}{2000}).

\bibitem[{\citenamefont{Takeda and K{\"u}bler}(1979)}]{Takeda79}
\bibinfo{author}{\bibfnamefont{T.}~\bibnamefont{Takeda}} \bibnamefont{and}
  \bibinfo{author}{\bibfnamefont{J.}~\bibnamefont{K{\"u}bler}},
  \bibinfo{journal}{J. Phys. F Met. Phys.} \textbf{\bibinfo{volume}{9}},
  \bibinfo{pages}{661} (\bibinfo{year}{1979}).

\bibitem[{\citenamefont{Jansen and Freeman}(1984)}]{Jansen84}
\bibinfo{author}{\bibfnamefont{H.~J.~F.} \bibnamefont{Jansen}}
  \bibnamefont{and} \bibinfo{author}{\bibfnamefont{A.~J.}
  \bibnamefont{Freeman}}, \bibinfo{journal}{Phys.\ Rev. B}
  \textbf{\bibinfo{volume}{30}}, \bibinfo{pages}{561} (\bibinfo{year}{1984}).

\bibitem[{\citenamefont{Vosko et~al.}(1980)\citenamefont{Vosko, Wilk, and
  Nusair}}]{VWN80}
\bibinfo{author}{\bibfnamefont{S.~H.} \bibnamefont{Vosko}},
  \bibinfo{author}{\bibfnamefont{L.}~\bibnamefont{Wilk}}, \bibnamefont{and}
  \bibinfo{author}{\bibfnamefont{M.}~\bibnamefont{Nusair}},
  \bibinfo{journal}{Can. J. Phys.} \textbf{\bibinfo{volume}{58}},
  \bibinfo{pages}{1200} (\bibinfo{year}{1980}).

\bibitem[{\citenamefont{Anisimov et~al.}(1997)\citenamefont{Anisimov,
  Aryasetiawan, and Lichtenstein}}]{Anisimov}
\bibinfo{author}{\bibfnamefont{V.~I.} \bibnamefont{Anisimov}},
  \bibinfo{author}{\bibfnamefont{F.}~\bibnamefont{Aryasetiawan}},
  \bibnamefont{and} \bibinfo{author}{\bibfnamefont{A.~I.}
  \bibnamefont{Lichtenstein}}, \bibinfo{journal}{J. Phys: Condens. Matter}
  \textbf{\bibinfo{volume}{9}}, \bibinfo{pages}{767} (\bibinfo{year}{1997}).

\bibitem[{\citenamefont{Solovyev et~al.}(1996)\citenamefont{Solovyev, Hamada,
  and Terakura}}]{Solovyev}
\bibinfo{author}{\bibfnamefont{I.}~\bibnamefont{Solovyev}},
  \bibinfo{author}{\bibfnamefont{N.}~\bibnamefont{Hamada}}, \bibnamefont{and}
  \bibinfo{author}{\bibfnamefont{K.}~\bibnamefont{Terakura}},
  \bibinfo{journal}{Phys.\ Rev. B} \textbf{\bibinfo{volume}{53}},
  \bibinfo{pages}{7158} (\bibinfo{year}{1996}).

\bibitem[{\citenamefont{Fernandez et~al.}(1998)\citenamefont{Fernandez,
  Vettier, de~Bergevin, Giles, and Neubeck}}]{Fernandez}
\bibinfo{author}{\bibfnamefont{V.}~\bibnamefont{Fernandez}},
  \bibinfo{author}{\bibfnamefont{C.}~\bibnamefont{Vettier}},
  \bibinfo{author}{\bibfnamefont{F.}~\bibnamefont{de~Bergevin}},
  \bibinfo{author}{\bibfnamefont{C.}~\bibnamefont{Giles}}, \bibnamefont{and}
  \bibinfo{author}{\bibfnamefont{W.}~\bibnamefont{Neubeck}},
  \bibinfo{journal}{Phys. Rev. B} \textbf{\bibinfo{volume}{57}},
  \bibinfo{pages}{7870} (\bibinfo{year}{1998}).

\bibitem[{\citenamefont{Sawatzky and Allen}(1984)}]{Sawatzky}
\bibinfo{author}{\bibfnamefont{G.~A.} \bibnamefont{Sawatzky}} \bibnamefont{and}
  \bibinfo{author}{\bibfnamefont{J.~W.} \bibnamefont{Allen}},
  \bibinfo{journal}{Phys. Rev. Lett.} \textbf{\bibinfo{volume}{53}},
  \bibinfo{pages}{2339} (\bibinfo{year}{1984}).

\bibitem[{\citenamefont{Cheetham and Hope}(1983)}]{Cheetham}
\bibinfo{author}{\bibfnamefont{A.~K.} \bibnamefont{Cheetham}} \bibnamefont{and}
  \bibinfo{author}{\bibfnamefont{D.~A.~O.} \bibnamefont{Hope}},
  \bibinfo{journal}{Phys.\ Rev. B} \textbf{\bibinfo{volume}{27}},
  \bibinfo{pages}{6964} (\bibinfo{year}{1983}).

\bibitem[{\citenamefont{Arola et~al.}(1997)\citenamefont{Arola, Strange, and
  Gyorffy}}]{Arola}
\bibinfo{author}{\bibfnamefont{E.}~\bibnamefont{Arola}},
  \bibinfo{author}{\bibfnamefont{P.}~\bibnamefont{Strange}}, \bibnamefont{and}
  \bibinfo{author}{\bibfnamefont{B.~L.} \bibnamefont{Gyorffy}},
  \bibinfo{journal}{Phys.\ Rev.\ B} \textbf{\bibinfo{volume}{55}},
  \bibinfo{pages}{472} (\bibinfo{year}{1997}).

\bibitem[{\citenamefont{Rennert}(1993)}]{Rennert1}
\bibinfo{author}{\bibfnamefont{P.}~\bibnamefont{Rennert}},
  \bibinfo{journal}{Phys.\ Rev.\ B} \textbf{\bibinfo{volume}{48}},
  \bibinfo{pages}{13559} (\bibinfo{year}{1993}).

\bibitem[{\citenamefont{Rennert}(1994)}]{Rennert2}
\bibinfo{author}{\bibfnamefont{P.}~\bibnamefont{Rennert}}, \bibinfo{journal}{J.
  Appl.\ Phys.} \textbf{\bibinfo{volume}{76}}, \bibinfo{pages}{6459}
  (\bibinfo{year}{1994}).

\bibitem[{\citenamefont{de~Bergevin and Brunel}(1981)}]{deBergevin1}
\bibinfo{author}{\bibfnamefont{F.}~\bibnamefont{de~Bergevin}} \bibnamefont{and}
  \bibinfo{author}{\bibfnamefont{M.}~\bibnamefont{Brunel}},
  \bibinfo{journal}{Acta Crystallogr., Sect. A: Cryst. Phys., Diffr., Theor.
  Gen. Crystallogr.} \textbf{\bibinfo{volume}{37}}, \bibinfo{pages}{324}
  (\bibinfo{year}{1981}).

\bibitem[{\citenamefont{Blume}(1985)}]{Blume1}
\bibinfo{author}{\bibfnamefont{M.}~\bibnamefont{Blume}}, \bibinfo{journal}{J.
  Appl. Phys.} \textbf{\bibinfo{volume}{57}}, \bibinfo{pages}{3615}
  (\bibinfo{year}{1985}).

\bibitem[{\citenamefont{Blume and Gibbs}(1988)}]{Blume2}
\bibinfo{author}{\bibfnamefont{M.}~\bibnamefont{Blume}} \bibnamefont{and}
  \bibinfo{author}{\bibfnamefont{D.}~\bibnamefont{Gibbs}},
  \bibinfo{journal}{Phys.\ Rev. B} \textbf{\bibinfo{volume}{37}},
  \bibinfo{pages}{1779} (\bibinfo{year}{1988}).

  
\bibitem[{\citenamefont{Coppens et~al.}(1999)\citenamefont{Coppens, Su, and
  Becker}}]{Coppens}
\bibinfo{author}{\bibfnamefont{P.}~\bibnamefont{Coppens}},
  \bibinfo{author}{\bibfnamefont{Z.}~\bibnamefont{Su}}, \bibnamefont{and}
  \bibinfo{author}{\bibfnamefont{P.~J.} \bibnamefont{Becker}},
  \emph{\bibinfo{title}{International Tables for Crystallography}}
  \bibnamefont{edited} \bibnamefont{by} \bibnamefont{A.~J.~C.} 
  \bibnamefont{Willson} \bibnamefont{and} \bibnamefont{E.} \bibnamefont{Prince}
  (\bibinfo{publisher}{Kluwer Academic}, \bibinfo{address}{BOSTON},
  \bibinfo{year}{1999}), vol.~\bibinfo{volume}{C}, chap.
  \bibinfo{chapter}{Analysis of charge and spin densities}, pp.
  \bibinfo{pages}{706--727}, \bibinfo{edition}{2nd} ed.

\bibitem[{\citenamefont{Blume}(1961)}]{Blume3}
\bibinfo{author}{\bibfnamefont{M.}~\bibnamefont{Blume}},
  \bibinfo{journal}{Phys.\ Rev.} \textbf{\bibinfo{volume}{124}},
  \bibinfo{pages}{96} (\bibinfo{year}{1961}).

\end{thebibliography}

\end{document}